\documentclass[a4paper, 11pt]{article}
\usepackage{graphicx}
\author{Clovis Jacinto de Matos\footnote{ESA-HQ, European Space Agency, 8-10 rue Mario Nikis,
75015 Paris, France, e-mail: Clovis.de.Matos@esa.int}}
\title{Physical Vacuum in Superconductors}

\begin{document}

\maketitle

\begin{abstract}
Although experiments carried out by Jain et al. showed that the
Cooper pairs obey the strong equivalence principle, The
measurement of the Cooper pairs inertial mass by Tate et al.
revealed an anomalous excess of mass. In the present paper we
interpret these experimental results in the framework of an
electromagnetic model of dark energy for the superconductors'
vacuum. We argue that this physical vacuum is associated with a
preferred frame. Ultimately from the conservation of energy for
Cooper pairs we derive a model for a variable vacuum speed of
light in the superconductors physical vacuum in relation with a
possible breaking of the weak equivalence principle for Cooper
pairs.
\end{abstract}

\section{Introduction}
In the present work we explore the consequences of the spontaneous
breaking of gauge invariance in superconductors on the principle
of equivalence for Cooper pairs. The breaking of gauge invariance
in superconductors makes the frame of the superconductor a
preferred frame. Thus it is natural to wonder if it is possible to
observe an absolute type of motion with respect to this frame.

Assuming that the idea of a physical vacuum in superconductors
defined by its energy momentum vector, with zero spatial component
and non-zero energy density, is correct, one deduces that this
type of vacuum is related to the superconductor's preferred frame.
Here we argue that below the superconductor's critical
temperature, the vacuum energy density in a superconductor
associated with a preferred frame corresponds to an
electromagnetic model of zero point energy contributing to a
vacuum energy density similar to the one from cosmological origin.
This model is intimately related with a discrete picture of
spacetime made by Minkowski's Diamond cells generated by the
continuous process of creation and annihilation of the Cooper
pairs. From this model we derive an index of refraction for the
vacuum speed of light in the superconductor, which appears to be
linked with a possible breaking of the weak equivalence principle
for the Cooper pairs. Experimental results supporting these
theoretical possibilities are reviewed.

In section 2, the experiment from Tate et al. measuring the Cooper
pairs inertial mass and the experiment from Jain et al. testing
the strong equivalence principle are analyzed with respect to a
possible breaking of the weak equivalence principle. In section 3
we present the consequences of the spontaneous breaking of gauge
invariance in superconductors with respect to their
electromagnetic and gravitational properties. In section 4 the
possibility of a variable vacuum speed of light in relation with
the physical implementation of a preferred frame in
superconductors is explored within a phenomenological perspective.
In section 5 we review the electromagnetic zero-point dark energy
model proposed by Beck, Mackey and the author, and introduce it as
a possible candidate for the vacuum energy in superconductors
required to physically implement locally a Lorentzian concept-type
of a physical ether. In section 6 we show that a variable speed of
light in superconductors reconciles the breaking of the weak
equivalence principle for cooper pairs with the law of energy
conservation. Finally we conclude with some avenues, which would
be worth to explore.
\section{Testing the Principle of Equivalence for Cooper pairs}
The Principle of General Covariance, which establishes the
independence of the laws of physics with respect to the physical
observer's reference frame, and which is at the foundation of
Einstein's theory of general relativity, can be formulated from
two different complementary phenomenological perspectives: The
Strong Equivalence Principle and the Weak Equivalence Principle.

The Strong Equivalence Principle states that locally the physical
effects of a uniform gravitational field are indistinguishable
from those due to an accelerating reference frame.

The Weak Equivalence Principle means the constancy of the ratio
between the inertial and the gravitational mass $m_i$ and $m_g$
respectively.
\begin{equation}
\frac{m_g} {m_i}=\iota=Cte\label{e0}
\end{equation}
This implies, in classical physics, that the possible motions in a
gravitational field are the same for different test particles.
Current experimental tests of the weak equivalence principle
\cite{Bassler} \cite{Smith}, indicate that the gravitational and
inertial masses of any physical body are equal to each other,
$m_g/m_i=\iota=1$, within a relative accuracy of the
E\"{o}tv\"{o}s-factor, $\eta(A,B)$ less than $5 \times 10^{-13}$.
\begin{equation}
\eta(A,B)=2 \frac {(m_g/m_i)_A - (m_g/m_i)_B}{(m_g/m_i)_A +
(m_g/m_i)_B}<5\times 10^{-13}\label{e1}
\end{equation}
The E\"{o}tv\"{o}s-factor is usually obtained from the measurement
of the differential acceleration, $\Delta a$, of two free falling
test bodies, $A$ and $B$.
\begin{equation}
\eta(A,B)=\frac{\Delta a}{g} \label{e2}
\end{equation}
where $g$ is the Earth's gravitational acceleration.

As argued by Anandan \cite{Anandan}, the principle of equivalence
cannot be demonstrated on a purely theoretical basis. Neither
classical or quantum physics can derive the equivalence principle
from more fundamental axioms. Thus it can only be justified by
experiment. In the following sections we present two important
experiments which bring relevant information about the validity of
the principle of general covariance for Cooper pairs in
superconductors.
\subsection{Anomalous Cooper pairs inertial mass excess}
In 1989 Cabrera and Tate \cite{Tate01, Tate02}, through the
measurement of the magnetic trapped flux originated by the London
moment, reported an anomalous Cooper pair inertial mass excess in
thin rotating Niobium superconductive rings:
\begin{equation}
\Delta m_i=m_i^*-m_i=94.147240(21)eV\label{e3}
\end{equation}
Here $m_i^*=1.000084(21)\times 2m_e=1.023426(21)MeV$ ($m_e$ being
the standard electron mass) is the experimentally measured Cooper
pair inertial mass (with an accuracy of 21 ppm), and
$m_i=0.999992\times2m_e=1.002331 MeV$ is the theoretically
expected Cooper pair inertial mass including relativistic
corrections.

This anomalous Cooper pair mass excess has not received, so far, a
satisfactory explanation in the framework of superconductor's
physics. If the gravitational mass of the Cooper pairs, $m_g$,
remains equal to the expected theoretical Cooper inertial mass,
$m_i=m_g=0.999992\times2m_e=1.002331 MeV$, Tate's experiment would
reveal that the Cooper pairs break the WEP with an
E\"{o}tv\"{o}s-factor, $\eta(E,T)=9.19\times 10^{-5}>>5\times
10^{-13}$, obtained from eq.(\ref{e1}) assuming the Experimental
$(E)$ and Theoretical $(T)$ ratios $m_g/m_i^*=0.999908$ and
$m_g/m_i=1$ respectively. The question is thus: Is an excess of
mass, similar to the one observed by Tate for the cooper pairs
inertial mass, also occurring for the cooper pair's gravitational
mass?
\subsection{Testing the Strong Equivalence Principle for Cooper Pairs}
In 1987 Jain et al carried out an experiment to probe the SEP for
Cooper pairs \cite{Jain}. The experiment consisted of two
Josephson junctions separated by a height $H=7.2 cm$, connected in
opposition by superconducting wires. In this experiment two
effects are competing to each other: On the one side the junctions
are coupled to microwave radiation from a common source, which
maintains a voltage difference of $2.35\times 10^{-21} V$ between
the two junctions by means of the gravitational red shift.
\begin{equation}
V_u=V_l(1-\frac{gH}{c^2})\label{d1}
\end{equation}
Where $V_u$ and $V_l$ are the electric potentials at the upper and
lower junctions respectively, and $g$ is the Earth gravitational
acceleration. On the other side the strong equivalence principle
predicts that the gravito-electromechanical potential $\bar{\mu}$
is constant along the connecting wires.
\begin{equation}
\bar{\mu}=\mu(1+\frac{m_g}{m_i}\frac{gH}{c^2})=Cte\label{d2}
\end{equation}
where $\mu$ is the electrochemical potential, which in general
will not be constant, $m_i$ and $m_g$ are the inertial and
gravitational Cooper pair masses. This implies that the potential
difference V between the superconducting wires varies with height
so that
\begin{equation}
V(z)=V(z=0)(1-\frac{m_g}{m_i}\frac{gH}{c^2})\label{d3}
\end{equation}
cumulating the two effects, eq.(\ref{d1}) and eq.(\ref{d3}), we
obtain the total loop emf:
\begin{equation}
\Delta
V=V_l(\frac{gH}{c^2}-\frac{m_g}{m_i}\frac{gH}{c^2})\label{d4}
\end{equation}
which is predicted to be zero on the basis of the strong
equivalence principle. Jain indeed measured the total emf to be
less than $1\times 10^{-22}$, consistent with the relativistic
prediction.

Therefore Jain et al. experiment tested the strong equivalence
principle for Cooper pairs, showing that using the Cooper pairs as
probe masses, we also reach the conclusion that the laboratory is
accelerating with respect to a local Minkowski spacetime. This
plainly justifies the curved spacetime description, which has been
well tested for neutral matter, to hold for Cooper pairs as well.
Since Jain's experiment tests the null result of eq.(\ref{d4}),
this experiment also demonstrated that the inertial and
gravitational mass of Cooper pairs are exactly equal to each other
within an accuracy of $4\%$:
\begin{equation}
\frac{m_i}{m_g}=1\pm0.04
\end{equation}
Unfortunately the accuracy of Jain's experiment is not good enough
to discard or confirm a difference between the inertial and the
gravitational mass of Cooper pairs of 21 ppm as reported by Tate
et al.
\subsection{Non detection of the Gravitomagnetic London Moment in Rotating Superconductors
 versus breaking of the weak equivalence principle for Cooper pairs}
As shown in \cite{Tajmar02}, Tate's experimental results could
also be interpreted as resulting from an additional
gravitomagnetic term in the Cooper pairs canonical momentum,
together with the assumption that the inertial and the
gravitational mass of the Cooper pairs, $m_i$ and $m_g$, remain
equal to their expected theoretical
values,$m_i=m_g=0.999992\times2m_e=1.002331 MeV$,
\begin{equation}
\vec {\pi}= m_i\vec v + e \vec A + m_g \vec {A_g}\label{e15}
\end{equation}
Where $\vec A$ is the magnetic vector potential, $v$ is the Cooper
pair velocity, and $\vec {A_g}$ is the gravitomagnetic vector
potential, whose rotational gives the gravitomagnetic field $\vec
B_g$:
\begin{equation}
\vec {B_g}=\nabla \times \vec{A_g}\label{e16}
\end{equation}
From Tate's measurements one can estimate the relative value of
the gravitomagnetic field, with respect to the superconductor's
angular velocity $\omega$, required to account for the anomalous
Cooper pair inertial mass excess:
\begin{equation}
\chi=\frac{(m_i^*-m_i)}{m_g}=\frac{B_g}{2 \omega}=9.2\times
10^{-5}\label{bie20}
\end{equation}
This equation clearly indicates that the interpretation of Tate's
experiment in terms of a gravitomagnetic term in the Cooper pairs
canonical momentum, only makes sense in the context of a breaking
of the weak equivalence principle for Cooper pairs with an
E\"{o}tv\"{o}s-factor, $\eta(E,T)=9.19\times 10^{-5}$, with $\eta$
and $\chi$ related to each other in the following manner:
\begin{equation}
\eta(E,T)=\frac{\chi}{1+\frac{\chi}{2}}\label{e21}
\end{equation}
In the case where $\chi<<1$, eq.(\ref{e21}) reduces simply to
\begin{equation}
\eta\sim\chi\label{eot}
\end{equation}
In the following we will refer to $\chi$ as the
"E\"{o}tv\"{o}s-factor $\chi$", and we refer to $\eta$ as the
"E\"{o}tv\"{o}s-factor $\eta$".

Therefore the E\"{o}tv\"{o}s-factor $\chi$ and $\eta$ can be
estimated not only through differential acceleration measurements
during free fall experiments but also from the measurement of the
gravitomagnetic Larmor theorem in rotating frames, with probe
masses located in strong gravitomagnetic fields:

We are left with two alternatives:

\begin{enumerate}

\item\label{i1} We can extrapolate Jain's experiment to the accuracy required to probe
Tate's results. This would imply that Cooper pairs do not violate
the strong and the weak equivalence principle, a null
E\"{o}tv\"{o}s-factor, $\eta(E,T)=0$, for Cooper pairs in
superconductors would then be expected. Therefore from
eq.(\ref{e21}) we would deduce that $\chi=0$. In this case the
Cooper pairs would still carry out an equal excess of inertial and
gravitational mass according to Tate's result, that needs to be
explained.

\item\label{i3} The Cooper pairs in Niobium break the weak equivalence principle with
an E\"{o}tv\"{o}s-factor, $\eta(E,T)=9.19\times 10^{-5}$. In this
case only the Cooper pairs inertial mass excess needs to be
explained.

\end{enumerate}

Although recent experiments from Tajmar et al. \cite{taj1}
involving rotating superconducting rings failed at detecting the
gravitomagnetic field appearing in eq.(\ref{bie20}), we see that
as indicated by eq.(\ref{e21}, the quotient $\chi=B_g/2\omega$ is
associated with the breaking of the weak equivalence principle for
Cooper pairs, and not with a gravitomagnetic analogue of the
magnetic London moment. Therefore the experiments carried out by
Tajmar et al. were not designed to detect the experimental effect
pertinent for the investigation of Tate's experimental results,
and which results from the correct phenomenological interpretation
of eq.(\ref{bie20}). Consequently the results from Tajmar et al.
experiments cannot provide us with relevant information to decide
which of the two alternatives above is the correct one.

In summary to investigate further Tate et al experimental results
we should not aim at detecting a gravitomagnetic type analog of
the London moment in rotating superconductors, but instead of this
we should aim at testing the weak equivalence principle for Cooper
pairs.

In the following we will try to demonstrate that the second
alternative above is the correct one. Thus we will demonstrate
that the Cooper pairs in superconductors could break the weak
equivalence principle.

\section{Spontaneous breaking of gauge symmetry in superconductors}
Superconductivity is a macroscopic quantum phenomena associated
with the formation of a condensate of electron pairs (called
Cooper pairs, or superelectrons) in the crystallin lattice of
certain solids below a critical temperature $T_c$, depending upon
the particular material. The Cooper pair condensate is described
by one single wavefunction $\psi(x^\mu)$ depending on the
spacetime coordinates $x^{\mu}=(x^0=ct, x^1=x, x^2=y, x^3=z)$.
\begin{equation}
\psi(x^\mu)=\alpha(x^\mu)e^{i\beta(x^\mu)}\label{1}
\end{equation}
where $n^*=\alpha^2$ represents the density of Cooper pairs, and
$\beta(x^\mu)$ is the phase of the wavefunction, both, $\alpha$
and $\beta$ are real valued functions. The average distance
between the two electrons in a Cooper pair is known as the
coherence length, $\xi_c$. Typically, the coherence length is
approximately 2 orders of magnitude larger than the interatomic
spacing of a solid, therefore Cooper pairs are not comparable with
tightly bound electron molecules. Instead, there are many other
electrons between those of a specific Cooper pair allowing for the
paired electrons to change partners on a time scale, $\tau$,
defined by Heisenberg's uncertainty principle,
\begin{equation}
\tau\leq\hbar/\Delta(0)\label{sp3}
\end{equation}
where $\hbar$ is Planck's constant.

According to Bardeen, Cooper, Schriefer (BCS) theory the binding
energy, $\Delta(0)$, between the electrons forming a Cooper pair
and the critical temperature, $T_c$, at $T=0$ in a given material
is:
\begin{equation}
\Delta(0)=1.76 k T_c\label{sp2}
\end{equation}
Where $k$ is Boltzmann constant. For low critical temperature
(conventional) superconductors $T_c\sim10K$, and $\Delta(0)\sim 1
meV$.

The 4-current density of Cooper pairs in a superconductor is:
\begin{equation}
j^\mu=\frac{2n^* e}{m_i^*}\Big(-\hbar
\frac{\partial\beta}{\partial x^\mu} - \frac {2e}{c} A_\mu
\Big)\label{2}
\end{equation}
where $m_i^*$ is the mass of the Cooper pair in the interior of
the superconductor, and $A_\mu(x^\mu)=(\phi, \vec A)$ is the
electromagnetic potential with time and space components being the
electric scalar potential $\phi$, and the magnetic vector
potential $\vec A$ respectively.

Superconductivity may be regarded fundamentally as being due to
the spontaneous breaking of the $U(1)$ gauge symmetry. This has
two fundamental consequences. first it makes \emph{the frame of
the superconductor a preferred frame $\Sigma$}. Second \emph{the
current density relative to the superconductor vanishes}. In the
limit of weak gravitational fields, the preferred frame is the
rest frame in which the superconductor 4-velocity is
\begin{equation}
t_\mu=(c_1,0,0,0,)\label{pref}
\end{equation}
In defining the 4-velocity $t\mu$ we have assumed a speed of light
$c_1$ which can be different or equal to the classical value $c$
to allow for the possibility of corrections to the predicted
relativistic effects. From eq.(\ref{2}), the vanishing of the
current density in the superconductor preferred frame means:
\begin{equation}
-\hbar \frac{\partial\beta}{\partial x^\mu} - \frac{2e}{c}
A_\mu=m_i^* t_\mu\label{3}
\end{equation}
The zeroth component of eq.(\ref{3}) leads to the electrochemical
potential $\mu$, which includes the rest-mass energy.
\begin{equation}
\mu=m_i^* c_1 c+ 2e A_0\label{4}
\end{equation}

In field theory the spontaneous breaking of gauge invariance leads
to massive photons via the Higgs mechanism. Massive photons can
also be understood as a consequence of the possibility of a
preferred frame in superconductors. In this case, in the
superconductor, the Maxwell equations transform to the so called
Maxwell-Proca equations, which are given by
\begin{eqnarray}
\nabla \vec{E}
&=&\frac{\rho^*}{\epsilon_0}-\frac{1}{\lambda_\gamma
^2}\phi \label{equ6}\\
\nabla \vec{B}&=&0 \label{equ7}\\
\nabla\times \vec{E}&=&-\dot{\vec{B}}\label{equ8}\\
\nabla\times\vec{B}&=&\mu_0 \rho^*
\vec{v_s}+\frac{1}{c^2}\dot{\vec
E}-\frac{1}{\lambda_\gamma^2}\vec{A} \label{equ9}
\end{eqnarray}
Where $\vec{E}$ is the electric field,
$\vec{B}=\bigtriangledown\times \vec A$ is the magnetic field,
$\epsilon_{0}$ is the vacuum electric permitivity, $\mu_{0}=1
/\epsilon_0  c^2$ is the vacuum magnetic permeability, $\phi$ is
the scalar electric potential, $\vec{A}$ is the magnetic vector
potential, $\rho^*=2en^*$ is the Cooper pairs fluid electric
density, $\vec{v_S}$ is the cooper pairs velocity, and
$\lambda_\gamma=\hbar/m_\gamma c$ is the photon's Compton
wavelength, which is equal to the London penetration depth
$\lambda_L=\sqrt{\frac{m}{\mu_0 n^* e}}$. Superconductor's
properties like the Meissner effect, and the London moment can be
derived from this set of equations \cite{de Matos01}.

The possibility of a preferred reference frame in superconductors
also allows to linearize Einstein Field Equations with a
cosmological constant \cite{clovis1}. This leads to the set of de
Sitter gravitoelectromagnetic equations, which include a massive
graviton.
\begin{eqnarray}
\nabla\vec g & = & + 3\pi G \rho -\frac{2}{3}\Lambda \varphi\label{ppequ22}\\
\nabla\vec{B_g} & = & 0 \label{ppequ23}\\
\nabla\times\vec g & = & -\frac{\partial \vec{B_g}}{\partial t} \label{ppequ24}\\
\nabla \times \vec B_g & = & -\frac{4 \pi G}{c^2} \rho \vec v_s +
\frac{1}{c^2} \frac{\partial \vec g}{\partial t} -\frac{2}{3}
\Lambda \vec A_g \label{ppequ25}
\end{eqnarray}
Where $\vec{g}$ is the gravitational field,
$\vec{B_g}=\bigtriangledown\times \vec A_g$ is the gravitomagnetic
field, $\epsilon_{0g}=1/3 \pi G$ is the vacuum gravitational
permittivity, $\mu_{0g}=4\pi G / c^2$ is the vacuum
gravitomagnetic permeability, $\varphi$ is the scalar
gravitational potential, $\vec{A_g}$ is the gravitomagnetic vector
potential, $\rho$ is the superconductor's physical vacuum mass
density, $\vec{v_S}$ is the cooper pairs velocity, $\Lambda$ is
the cosmological constant, and $\lambda_g$  is the Compton
wavelength of the massive graviton.
\begin{equation}
\frac {1} {\lambda_g^2}=\Big( \frac {m_g c} {\hbar} \Big )^2=
\frac {2\Lambda}  {3}\label{graviton}
\end{equation}
The de Sitter gravitoelectromagnetic set of equations is only
valid locally, at the origin of the preferred reference frame
attached to the superconductor, and since repulsive gravitational
fields are predicted by eq.(\ref{ppequ22}) they only apply to the
energy density of the superconductor's physical vacuum. Taking the
rotational of eq.(\ref{ppequ25}), and solving the resulting
differential equation for the 1-dimensional case of a
superconducting ring, rotating with angular velocity $\omega$, we
find the gravitomagnetic Larmor theorem \cite{Mashhoon} for Cooper
pairs:
\begin{equation}
B_g= 2 \omega\mu_{0g} \rho \lambda^2_g \label{equ17}
\end{equation}

The interpretation of a non-zero cosmological constant in Einstein
field equations as the physical possibility of privileged
coordinate frames without breaking the strong equivalence
principle, was already debated by Rayski in \cite{Rayski}.
Although the strong equivalence principle is the backbone of the
principle of general covariance, the gravitational analogue of
electromagnetic gauge invariance is the weak equivalence principle
\cite{Weinberg}. Therefore the spontaneous breaking of gauge
invariance in superconductors would appear together with the
breaking of the weak equivalence principle for Cooper pairs. Since
the breaking of gauge invariance is affecting the photon rest
mass, is the existence of a preferred frame attached to a
superconductor affecting also the constant value of the speed of
light in vacuum?
\section{Vacuum speed of light in superconductors}
Starting from Consoli and Costanzo idea that the physical vacuum
might be defined by its energy momentum vector, with zero spatial
component and non-zero energy density for the time component
\cite{Consoli}\cite{consoli2}, one fixes the 4-velocity of the
vacuum medium in a similar manner as we defined above the
preferred frame attached to the superconductor, eq.(\ref{pref})
\begin{equation}
t'^\mu(S')\equiv(c',0,0,0)\label{con1}
\end{equation}
where the vacuum speed of light $c'$ is not necessarily equal to
its classical value $c$ or to the superconductor's preferred frame
vacuum speed of light $c_1$.

An observer attached to the superconductor preferred frame,
$\Sigma$, would witness locally two possible vacuum speeds of
light $c$ and $c'$, this will affect locally the laws of special
relativity for this observer. The diagonal form of the interval of
universe with respect to $\Sigma$ before the material becomes
superconductor is the standard one:
\begin{equation}
ds^2=c^2dt^2-dl^2\label{spec1}
\end{equation}
When the material becomes superconductor the same observer in
sigma will observe a different diagonal form for the interval of
universe $ds$ that will change to $ds'$ according to the
superconductor vacuum refractive index $N_{vacuum}$:
\begin{equation}
\Big( \frac{ds}{ds'}\Big )^2=N_{vacuum}=\frac{c}{c'}\label{spec2}
\end{equation}
Therefore the Minkowski interval of universe in the
superconducting state will be:
\begin{equation}
ds^2=\frac{c}{c'} \Big[ (2cc'-c^2)dt-dl^2\Big]\label{spec3}
\end{equation}
From this interval we deduce that the effective Lorentzian speed
of light that will set the usual relativistic effects for the
superconductor will be:
\begin{equation}
c_{eff}^2=2cc'-c^2=2c_g^2-c^2\label{spec4}
\end{equation}
where $c_g=cc'$ is the geometric mean between $c$ and $c'$.

The time dilatation relative to the superconductor will be
expressed in function of this effective speed of light.
\begin{equation}
dt=\frac{d\tau}{\sqrt{1-\frac{v^2}{c_{eff}^2}}}\label{spec5}
\end{equation}
Accordingly the law of length contraction will be.
\begin{equation}
dl=dl_0 \sqrt{1-\frac{v^2}{c_{eff}^2}}\label{spec6}
\end{equation}
The increase of mass-energy with velocity will be:
\begin{equation}
E=\frac{m_0
c_{eff}^2}{\sqrt{1-\frac{v^2}{c_{eff}^2}}}\label{spec7}
\end{equation}

The spacetime metric is only diagonal with respect to $\Sigma$.
The relative velocity $v$ between an observer $S'$, \emph{located
outside the volume of the superconductor}, and $\Sigma$ introduces
off diagonal elements $g_{0i}$ in the metric relative to $S'$. By
starting with the diagonal and isotropic form eq.(\ref{spec3}),
one obtains
\begin{equation}
g_{0i}\sim2 (N_{vacuum} - 1) \frac{v_i}{2c'-c}\label{con1a}
\end{equation}
Since the components of the gravitomagnetic vector potential $\vec
A_g=(A_{g1}, A_{g2}, A_{g3})$ results from the spacetime metric
components $g_{0i}$
\begin{equation}
A_{gi}=-cg_{0i}\label{con5}
\end{equation}
And since the gravitomagnetic field $\vec B_g$ is obtained from
the rotational of the gravitomagnetic vector potential
\begin{equation}
\vec B_g=\nabla\times\vec A_g\label{con6}
\end{equation}
Multiplying eq.(\ref{con1a}) by $2c'-c$ and taking the rotational
of both sides of the equation we obtain
\begin{equation}
-2\frac{B'_g}{2\omega}+\frac{B_g}{2 \omega}\sim 2
(N_{vacuum}-1)\label{con7}
\end{equation}
where $B'_g=\nabla\times (-c'g_{0i})$ is the gravitomagnetic field
existing in the superconductor due to the non-classical vacuum. At
this stage of our rational the non-classical E\"{o}tv\"{o}s-factor
$\chi'=B'_g/2\omega$ is an unknown quantity.

We reach the conclusion that the refractive index of the vacuum
depends on the E\"{o}tv\"{o}s-factor, $\chi=B_g/2\omega$, and vis
versa.
\begin{equation}
\chi'+ \frac{\chi}{2}\sim N_{vacuum}-1\label{con8}
\end{equation}
Therefore a preferred frame resulting from a breaking of gauge
invariance in a superconductor would lead simultaneously to a
breaking of the weak equivalence principle for Cooper pairs and to
a variable speed of light in the superconductor's vacuum. The
experimental detection of one of this effects should imply the
existence of the other. The next question that naturally arise is
about the physical nature of the vacuum in the superconductor.
\section{Electromagnetic zero-point dark energy and discrete
spacetime in superconductors}

A non-vanishing cosmological constant can be interpreted in terms
of a non-vanishing vacuum energy density, $\rho_0$.
\begin{equation}
\rho_0=\frac{c^4}{8 \pi G} \Lambda\sim 10^{-29} g~cm^{-3} \simeq
3.88 e V/mm^3  \label{e23}
\end{equation}
where $\Lambda=1.29\times 10^{-52} [m^{-2}]$ is the cosmological
constant \cite{Spergel}. The cosmological constant is a good
candidate to account for the dark-energy resulting from the latest
cosmological observations reporting an accelerated expansion of
the universe according to the following equation of state:
$\rho_0=-p$, where $p$ is the pressure. In \cite{Beck4} Beck and
Mackey have developed an electromagnetic model of vacuum energy in
superconductors. This model is based on bosonic vacuum
fluctuations creating a small amount of vacuum energy density. One
assumes that in the superconducting phase the photons, with zero
point energy $\varepsilon=\frac{1}{2} h \nu$, contribute to a
vacuum energy density, $\rho^*$, similar to the vacuum energy
density resulting from the cosmological constant, eq.(\ref{e23}),
and to which we will often refer as "electromagnetic zero-point
dark energy". This vacuum energy density depends on a certain
frequency cutoff $\nu_c$.
\begin{equation}
\rho^*=\frac{1}{2}\frac{\pi h}{c^3} \nu_c^4 \label{e24}
\end{equation}
In \cite{Beck4} the formal attribution of a temperature $T$ to the
graviphotons is done by comparing their zeropoint energy with the
energy of ordinary photons in a bath at temperature $T$:
\begin{equation}
\frac{1}{2} h\nu=\frac{h\nu}{e^{\frac{h\nu}{kT}}-1}\label{e25}
\end{equation}
This condition is equivalent to
\begin{equation}
h\nu=\ln3kT\label{e26}
\end{equation}
Substituting the critical transition temperature $T_c$ specific to
a given superconductive material in Eq.(\ref{e26}), we can
calculate the critical frequency characteristic for this material:
\begin{equation}
\nu_c=\ln3 \frac{kT_c}{h}\label{lie27}
\end{equation}
For example, for Niobium with $T_c=9.25$K we get $\nu_c=0.212$
THz, which when used as a cutoff frequency in eq.(\ref{e24}) leads
to a vacuum energy $\rho^*=0.49 meV/mm^3$ inside the
superconductor. Substitution of eq.(\ref{lie27}) in eq.(\ref{e24})
leads to the law defining the density of electromagnetic
zero-point dark energy in function of the superconductor's
critical temperature, $T_c$.
\begin{equation}
\rho^*=\frac{\pi \ln^4 3}{2}\frac{k^4}{(ch)^3}T_c^4\label{e12}
\end{equation}
If the zero-point electromagnetic dark energy has the same
equation of state has the cosmological dark energy, $\rho_0$, then
$\rho^*$ should exert a tiny negative pressure on the
superconductor. By substitution of eq.(\ref{graviton}),
eq.({\ref{e23}) and eq.(\ref{e12}) in eq.(\ref{equ17}) we obtain
the E\"{o}tv\"{o}s-factor, $\chi$, quantifying the level of
breaking of the WEP for the Cooper pairs in a given superconductor
in function of the superconductor's critical temperature, $T_c$.
\begin{equation}
\chi=\frac{3}{2}\frac{\rho^*}{\rho_0}=\frac{3\ln^4 3}{8
\pi}\frac{k^4G}{c^7\hbar^3 \Lambda} T_c^4\label{12}.
\end{equation}
Remarkably, this equation connects the five fundamental constants
of nature $k,G,c,\hbar, \Lambda$ with measurable quantities in a
superconductor, $\chi$ and $T_c$.

We may define a Planck-Einstein temperature scale $T_{PE}$ as
\begin{equation}
T_{PE}=\frac{1}{k}\Bigg(\frac{c^7\hbar^3
\Lambda}{G}\Bigg)^{1/4}=60.71 K. \label{13}
\end{equation}
and the corresponding Planck-Einstein length
\begin{equation}
l_{PE}=\frac{\hbar}{M_{PE}c}=\Bigg(\frac{\hbar G}{c^3
\Lambda}\Bigg)^{1/4}=0.037[mm]\label{ex19}
\end{equation}
which is of the same order of magnitude as the Cooper pairs
coherence length $\xi_c$. Eq.(\ref{12}) can then be written as
\cite{dematos}
\begin{equation}
\chi=\frac{3\ln^4 3}{8
\pi}\Bigg(\frac{T_c}{T_{PE}}\Bigg)^4.\label{14}
\end{equation}
Substituting the critical transition temperature of Niobium,
$T_c=9.25K$, in Eq.(\ref{14}) we find the following
E\"{o}tv\"{o}s-factor $\chi$ for superconductive Niobium:
\begin{equation}
\chi=9.35\times10^{-5}\label{15}
\end{equation}
The above theoretically predicted value is close to the measured
value in Cabrera and Tate's experiment, Eq.(\ref{bie20}):
\begin{equation}
\chi=\frac{(m_i^*-m_i)}{m_g}=\frac{B_g}{2 \omega}=9.2\times
10^{-5}\label{e20}
\end{equation}

In summary we found that by extending the initial Beck and Mackey
model of electromagnetic dark energy to superconductor's critical
temperatures, different from the critical temperature associated
with the cosmological constant cutoff frequency, we predict an
E\"{o}tv\"{o}s-factor $\chi$ for Cooper pairs in superconducting
Niobium very close to the one estimated from Tate et al
experiment. This is a very encouraging result with respect to our
interpretation, in section 2.3, of the Cooper pairs inertial mass
excess being related with a breaking of the weak equivalence
principle for these particles.

The attempt to resolve the inverse cosmological constant problem
in \cite{Beck01}, where formally the cosmological constant comes
out 120 orders of magnitude too small, leads to assume that the
spacetime volume filled by Cooper pairs in a superconductor is
made of Planck-Einstein cells having a 4-volume, $l_{PE}^4$, which
will statistically fluctuate according to:
\begin{equation}
\Delta V\sim\sqrt V l_{PE}^2\label{ble25}
\end{equation}
Since the density of vacuum energy associated with the
cosmological constant $\rho_{0}$, is canonically conjugated with
the universe four-volume $V$, we can formulate the following
4-dimensional Heisenberg uncertainty principle for the cosmos.
\begin{equation}
\Delta \rho_{0} \Delta V \sim \hbar c \label{ble26}
\end{equation}
By substitution of $\rho_{0}$ and $\Delta V$ in eq.(\ref{ble26}),
by the electromagnetic zero-point dark energy density,
eq.(\ref{e12}), and the SC's four volume fluctuations,
eq.(\ref{ble25}) respectively, we deduce that the superconductor's
E\"{o}tv\"{o}s-factor $\chi$ statistically fluctuates according to
the quantum fluctuations of the Cooper pairs's discrete spacetime
volume:
\begin{equation}
\chi \sqrt{V} \sim \frac{\pi ^2}{3} l^2_{PE}\label{e27}
\end{equation}

A discrete structure of the spacetime volume spanned by the Cooper
pairs \cite{demat}, in terms of Planck-Einstein cells, with volume
fluctuating around the Planck-Einstein volume, $l_{PE}^4$, can
find a fundamental basis in relation with the Unruh effect for
finite lifetime inertial observers \cite{rovelli}. For an observer
with constant acceleration and infinite lifetime, the vacuum state
of a suitable quantum field theory in Minkowski spacetime appears
as a thermal equilibrium state with temperature
\begin{equation}
T_U=\frac{\hbar a}{2 \pi k c}\label{u1}
\end{equation}
This result can be derived from the integration along the
worldline of the observer of the interaction term between a
detector and the vacuum. Martinetti and Rovelli \cite{rovelli}
argued that for an inertial observer (i.e. with zero acceleration)
with finite-proper lifetime $\tau$, the finite Minkowski spacetime
region with 4-volume $(c\tau)^4$, called a Diamond, appears as a
thermal equilibrium state with temperature
\begin{equation}
T_D=\frac{2\hbar}{\pi k \tau}\label{u2}
\end{equation}
$T_D$ is designated as the \emph{Diamond temperature} associated
with the spacetime Diamond shaped region spanned by the inertial
observer in Minkowski spacetime. Let us assimilate a Cooper pair,
in a given superconductor, to an inertial observer with (periodic)
proper lifetime $\tau$ given by the substitution of eq.(\ref{sp2})
in eq.(\ref{sp3})
\begin{equation}
\tau\leq\frac{\hbar} {1.76 k T_c}\sim t_{PE}=\frac
{l_{PE}}{c}\label{u3}
\end{equation}
Substitution of eq.(\ref{u3}) in eq.(\ref{u2}) gives a lower limit
for the Diamond temperature of Cooper pairs.
\begin{equation}
T_D\sim\frac{3.52}{\pi}T_c=1.12 T_c\label{u4}
\end{equation}
Substituting eq.(\ref{u4}) in the expression of the
electromagnetic zero-point dark energy quanta, eq.(\ref{e26}) we
get
\begin{equation}
h\nu_c=\frac{\ln 3}{1.12}k T_D = 0.98 k T_D \sim k T_D\label{u5}
\end{equation}
This equation means that we can assimilate a quanta of
electromagnetic zero-point dark energy in a superconductor with a
Planck-Einstein sized, $l_{PE}^4$, Cooper pair Diamond cell.

Since the gravitational dark energy quantum condensate is related
with zero-point fluctuations, the physical model of the
superconductor's vacuum, presented above violates the principle of
energy conservation. This fact is also expected from a breaking of
the weak equivalence principle for Cooper pairs, which also leads
to non-conservation of energy-momentum. Would a simultaneous
breaking of the weak equivalence principle for Cooper pairs,
together with a variable speed of light in the superconductor
reconcile the electromagnetic zero-point dark energy model in
superconductors with the law of energy conservation?

\section{Variable speed of light and breaking of the weak equivalence principle in superconductors}
In figure (\ref{fig2}) we plotted the value of the
E\"{o}tv\"{o}s-factor $\chi$, eq.(\ref{14}), in function of the
superconductor's critical temperature. From this curve we see that
the difference between the inertial and the gravitational mass of
the Cooper pairs should continuously increase with the increase of
the superconductor's critical temperature. If the speed of light
remains constant the difference between the electrons rest mass
energy before forming the Cooper pair and the rest mass energy of
a Cooper pair should progressively increase with the
superconductors critical temperature. Therefore a breaking of the
weak equivalence principle for Cooper pairs with a constant speed
of light in the superconductors would violate the law of energy
conservation. Would a variable speed of light predicted by
eq.(\ref{con8}) compensate for the Cooper pairs inertial mass
increase in a way that would preserve energy conservation?

In the following we will impose that the rest mass energy of the
Cooper pairs should be conserved independently of the breaking of
the weak equivalence principle. In other words the Cooper pairs
theoretical and experimental (as measured by Tate) rest mass
energy should be equal to each other. Setting the magnetic term in
the electrochemical potential of Cooper pairs in eq.(\ref{4})
equal to zero, and defining the superconductor 4-velocity
$t_\mu(c_1,0,0,0)$, if the Cooper pairs rest mass energy is
conserved we should have
\begin{equation}
m_i c^2= cc_1m_i^*\label{dis1}
\end{equation}
where $m_i$ is the Cooper pairs theoretical inertial mass and
$m^*_i$ is the experimental Cooper pairs mass measured by Tate et
al. Setting $v=0$ in eq.(\ref{spec7}) we obtain the proper rest
mass energy of Cooper pairs in the Cooper pairs rest frame, i.e.,
in the superconductor preferred frame $\Sigma$.
\begin{equation}
m_i^* c c_1 =m_i^*c(2c'-c)\label{dis2}
\end{equation}
Note that from this equation we deduce that the speed of light
associated with the superconductotrs preferred frame is:
\begin{equation}
c_1=2c'-c.\label{dis2bis}
\end{equation}

Setting eq.(\ref{dis1}) equal to eq.(\ref{dis2}) we get
\begin{equation}
m_ic^2=m_i^*(2cc'-c^2)\label{dis3}
\end{equation}
From eq.(\ref{dis3}) we deduce the superconductor's vacuum
refractive index in function of the E\"{o}tv\"{o}s-factor $\chi$.
\begin{equation}
N_{vacuum}-1=\frac{\chi}{\chi+2}\label{dis4}
\end{equation}
which we can also express in function of the E\"{o}tv\"{o}s-factor
$\eta$ using eq.(\ref{e21})
\begin{equation}
N_{vacuum}-1=\frac{\eta}{2}\label{dis4bis}
\end{equation}
Substituting eq.(\ref{14}) in eq.(\ref{dis4}) we obtain the
variation of the superconductor's vacuum refractive index with
respect to the superconductor's critical temperature $T_c$.
\begin{equation}
N_{vacuum}-1=\frac{1}{1+\frac{16\pi}{3 \ln^4
3}\Big(\frac{T_{PE}}{T_c}\Big)^4} \label{distc}
\end{equation}

From eq.(\ref{dis4}) we deduce the speed of light $c'$ associated
with the superconductors vacuum, in function of the
E\"{o}tv\"{o}s-factor $\chi$.
\begin{equation}
c'=c\frac{1+\frac{\chi}{2}}{1+\chi}\label{dis5}
\end{equation}
which we can also express in function of the E\"{o}tv\"{o}s-factor
$\eta$
\begin{equation}
c'=\frac{2c}{\eta+2}\label{diset}
\end{equation}
Solving eq.(\ref{diset}) with respect to $c'$ we obtain the
variation of the superconductor's vacuum velocity $c'$ with
respect to the superconductor's critical temperature $T_c$.
\begin{equation}
c'=c\Big[ \Big( 1+\frac{16 \pi}{3\ln^4
3}\Big(\frac{T_{PE}}{T_c}\Big)^4\Big)^{-1}+1 \Big]^{-1}\label{cTc}
\end{equation}
As $T_c$ tends to infinity $c'$ tends to $c/2$

Setting eq.(\ref{dis4}) equal to eq.(\ref{con8}) we deduce the
non-classical E\"{o}tv\"{o}s-factor $\chi'=B'_g/2\omega$
\begin{equation}
\chi'=\frac{\chi^2}{2(\chi+2)}\label{dis6}
\end{equation}
We see that $\chi'$ is a second order term with respect to $\chi$.

Substituting eq.(\ref{dis5}) in the equation of the effective
Lorentzian speed of light eq.(\ref{spec4}), we deduce this speed
in function of the E\"{o}tv\"{o}s-factor $\chi$.
\begin{equation}
c_{eff}=\frac{c}{\sqrt{1+\chi}}\label{dis7}
\end{equation}
In a similar manner as above we can express $c_eff$ in function of
the E\"{o}tv\"{o}s-factor $\eta$
\begin{equation}
c_{eff}=c\Big(\frac{2-\eta}{2+\eta}\Big)^{1/2}\label{dis7eta}
\end{equation}
Substituting eq.(\ref{14}) in eq.(\ref{dis7}) we obtain the
variation of $c_{eff}$ with respect to the superconductor's
critical temperature $T_c$.
\begin{equation}
c_{eff}=\frac{c}{\sqrt{1+\frac{3\ln^4
3}{8\pi}\Big(\frac{T_c}{T_{PE}}\Big)^4}}\label{ceff}
\end{equation}
As $T_c$ tends to infinity $c_{eff}$ tends to zero.

Substituting eq.(\ref{cTc}) in eq.(\ref{dis2bis}) we obtain the
speed of light $c_1$ with respect to the superconductor's
preferred frame $\Sigma$ in function of the supercondutor's
critical temperature $T_c$.
\begin{equation}
c_1=c\Big(2\Big[ \Big( 1+\frac{16 \pi}{3\ln^4
3}\Big(\frac{T_{PE}}{T_c}\Big)^4\Big)^{-1}+1
\Big]^{-1}-1\Big)\label{c111}
\end{equation}

Figure (\ref{fig3}) displays the plots of $c'$ eq.(\ref{cTc}),
$c_{eff}$ eq.(\ref{ceff}) and $c_1$ eq.(\ref{c111}).

\section{Conclusions}
In conclusion the Cooper pairs bosonic condensate in
superconductors seems to generate a physical medium similar to a
superconducing "ether" constituted by a discrete set of
Minkowski's Diamond cells, each diamond cell representing a quanta
of electromagnetic zero-point dark energy in the superconductor. A
variable vacuum speed of light in superconductors would appear in
relation with a breaking of the weak equivalence principle,
ultimately resulting from a spontaneous breaking of gauge
invariance leading to a preferred frame in superconductors. We
have seen that a variable vacuum speed of light in superconductors
would allow to reconcile a breaking of the weak equivalence
principle for Cooper pairs with the principle of energy
conservation.

Since the breaking of the weak equivalence principle is only
affecting the Cooper pairs it is very difficult to detect this
effect with macroscopic superconducting samples, because the
Cooper pairs are only contributing marginally to the total
inertial and gravitational mass of the sample. Although this is a
major difficulty to test the weak equivalence principle for Cooper
pairs, the results presented in the present work lead to recommend
to carry out new versions of Jain and Tate experiments with
improved accuracy in order to measure the inertial and the
gravitational mass of Cooper pairs and therefore to measure the
E\"{o}tv\"{o}s factor for Cooper pairs with improved accuracy.
Streamlined with this recommendation it would be important to
carry out the non-null version of Jain's experiment suggested by
Anandan in \cite{Anandan}.

The possibility of two different vacuum speeds of light $c$ and
$c'$, with $c/2\leq c'\leq c$, for an observer attached to the
superconductor's preferred frame affects locally the classical
diagonal form of the Minkowski's spacetime metric relative to the
superconductor's preferred frame. We have shown that the effective
speed of light $c_{eff}$ setting the relativistic effects for the
superconductor with respect to an observer external to the
superconductor, is a function of the geometric mean between the
two possible speeds of light eq.(\ref{spec4}),
$c_{eff}^2=2cc'-c^2=2c_g^2-c^2<c$. Therefore we cannot accelerate
a superconductor until it approaches asymptotically the classical
vacuum speed of light $c$. \emph{Contrary to ordinary matter}, the
speed of a superconductor can only approach the effective speed of
light $c_eff<c$ depending on the fourth power of the
superconductor's critical temperature. to illustrate this result
let us think about an hypothetical superconductor with an infinite
critical temperature, according to the theory presented here it
would be impossible to communicate a relative speed different from
zero to the superconductor. so to speak, this superconductor would
be in an absolute state of rest.

It is interesting to note that the main ether drift experiments
referenced by Consoli in \cite{consoli2} contain optical
superconducting cavities. A possible question raised by the
present work is: Are the observed anisotropy in the speed of light
in Hermann \cite{Hermann} and Muller \cite{Muller} experiments due
to the presence of superconducting optical cavities in their
experiments?

\section{Acknowledgements}
The author would like to acknowledge Dr. Martin Tajmar,  and Prof.
Christian Beck, for continuous support and regular exchange of
information.

\begin{figure}[h!]
\begin{center}
\includegraphics[scale=0.6]{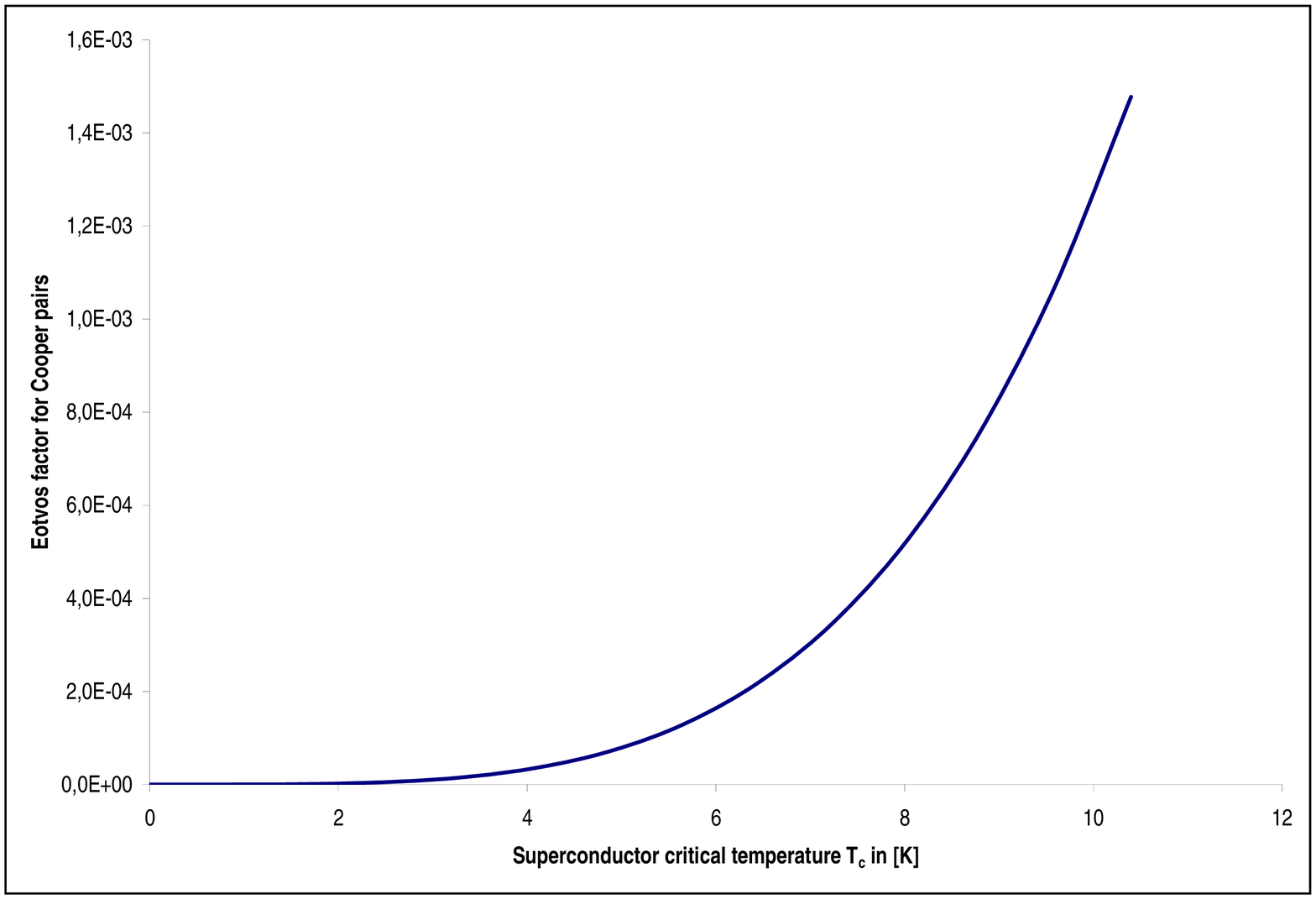}
\caption{\label{fig2} E\"{o}tv\"{o}s-factor $\chi$, quantifying
the level of breaking of the weak equivalence principle for Cooper
pairs, in function of the superconductor's critical temperature
$T_c$, eq.(\ref{14}).}
\end{center}
\end{figure}

\begin{figure}[h!]
\begin{center}
\includegraphics[scale=0.6]{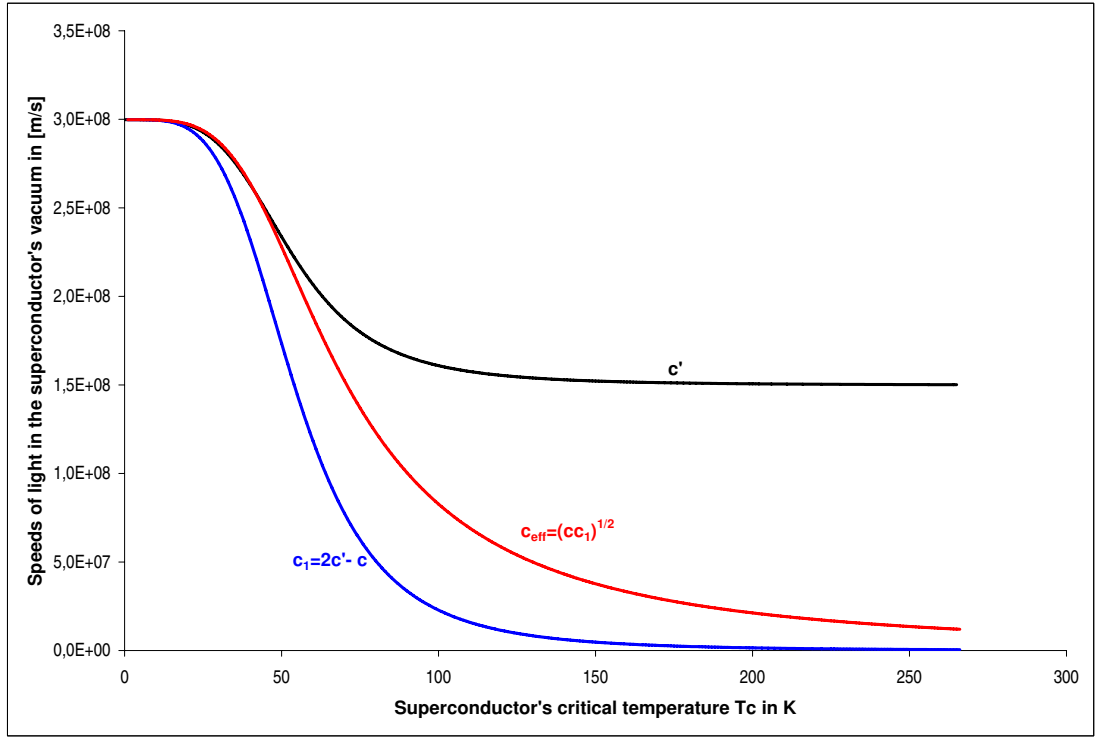}
\caption{\label{fig3} Decrease of the speed of light associated
with the superconductor electromagnetic dark energy vacuum medium,
$c'$ eq.(\ref{cTc}); decrease of the speed of light in the
superconductor's preferred frame $c_1$ eq.(\ref{c111}); and
decrease of the effective Lorentzian speed of light $c_{eff}$
eq.(\ref{ceff}) defining the usual relativistic effects for the
superconductor; in function of the superconductor's critical
temperature.}
\end{center}
\end{figure}

\end{document}